\documentstyle[11pt,twoside,fleqn,espcrc2]{article}

\newcommand{\AmS}{{\protect\the\textfont2
  A\kern-.1667em\lower.5ex\hbox{M}\kern-.125emS}}

\title{Future Reactor Neutrino Oscillation Experiments at Krasnoyarsk}

\author{L. Mikaelyan\address{Kurchatov Institute, Moscow, Russia}%
}
\begin{document}

\begin{abstract}
Recent studies of atmospheric neutrinos and the results from CHOOZ and Palo-Verde experiment 
call for new and more sensitive searches for neutrino oscillations at reactors. The main goal of the 
project considered here is to look for very small mixing angle oscillations of electron neutrinos in the 
atmospheric neutrino mass parameter region around ${\Delta}m^{2} \sim 3\times10^{-3} eV^{2}$ and to define 
the element $U_{e3}$ 
of the neutrino mixing matrix ($U_{e3}$ is the contribution of the mass-3 state to the electron neutrino flavor 
state). The practical goal of the project is to decrease, relative to the CHOOZ, the statistic and 
systematic errors as much as possible. \ To achieve this we plan to use two identical antineutrino 
detectors each with a $\sim$ 50-ton liquid scintillator target located at $\sim$1100 m and $\sim$250 m from the 
underground reactor ($\sim$ 600 mwe). \ Much attention is given to the detector calibration and monitoring 
procedures. As a first step we consider two much smaller pilot detectors each of $\sim$ a 3 ton target mass 
stationed at $\sim$20 m and 35$-$60 m from the reactor. \ The goals of this first stage are: (i) to accumulate 
necessary experience and (ii) to investigate with electron neutrinos the LSND mass parameter region.
\end{abstract}

\maketitle

\section{INTRODUCTION}

The $\sim$ 1 km baseline reactor experiments CHOOZ [1] and Palo Verde [2] have searched for 
electron antineutrino disappearance in the atmospheric mass parameter region 
(${\Delta}m^{2} \sim 3\times10^{-3} eV^{2}$) using the ${\bar{\nu}}_{e} + p \rightarrow e^{+} + n$
process as a detection reaction. No signs of oscillations have been found:
\begin{eqnarray}
\quad \sin^{2}2\theta_{CHOOZ} \le O.1	 \nonumber \\
\quad (at \quad {\Delta}m^{2}=3\times 10^{-3} eV^{2}).
\end{eqnarray}

The Super Kamiokande data reported at this Conference [3] confirm the evidence in favor 
of intensive ($\sin^{2}2\theta_{SK}\approx 1)\quad {\nu}_{\mu} \rightarrow {\nu}_{x}(x \ne e)$ 
transitions. In the three active neutrino mixing model \ considered here \quad ${\nu}_{\mu} = {\nu}_{\tau}$. 

The negative results of the CHOOZ experiment has important positive meaning that the 
contribution $U_{e3} (=\sin{\theta}_{13})$ of mass-3 eigenstate to the ${\nu}_e$ flavor state
is not large:
\begin{equation}
\qquad U_{e3} \le 2.3\times 10^{-2}
\end{equation}
(We remember that $\sin^{2}2{\theta}_{CHOOZ}$ = $4{U_{e3}}^{2} (1 - {U_{e3}}^{2})\approx 4 {U_{e3}}^{2}$, \
 see for example [4]).

In this report we consider a new reactor experiment Kr2Det aimed for much more sensitive 
search of neutrino oscillations in the ${\Delta}m^{2}_{atm}$ mass parameter region. The main physical goals of the 
search are:
\begin{itemize}
\item To obtain new information on the electron neutrino mass structure ($U_{e3}$),
\item To provide normalization for future \\ long baseline experiments at accelerators,
\item To achieve better understanding of the role ${\nu}_e$ can play in the atmospheric neutrino anomaly.
\end{itemize}

It is interesting to note that measurement of $U_{e3}$ can help to choose between possible solar neutrino 
oscillation solutions [5]

Other physical potentialities of the project (test of the LSND results, search for the sterile 
neutrinos) are considered elsewhere. 
In the next Section we briefly review factors, which limited the sensitivity of the CHOOZ results 
and then go to the Kr2Det project.

\vspace{10mm}

\section{THE CHOOZ EXPERIMENT}

The CHOOZ detector was built in an underground gallery (300 mwe) at a distance of 
about 1 km from two PWR type reactors of total rated power 8.5 GW (th). \ Using $e^{+}, n$ delayed 
coincidence technique total about 2500 neutrino interactions were detected in the 5-ton liquid 
scintillator target. The neutrino and background detection rates, $N_{\nu}$ and $N_{BKG}$ , were measured to 
be (typically) 12 (d$^{-1}$) and 1.6 (d$^{-1}$). \quad The ratio $R_{meas/calc}$ of measured to calculated for 
no-oscillation case neutrino rates was found to be 
\begin{eqnarray}
\quad R_{meas/calc}= 1.01 \pm 2.8\%(stat) \nonumber \\
\quad \pm 2.7\%(syst),
\end{eqnarray}

The ratio of measured to expected positron spectra $S_{meas}/S_{expected}$ is presented 
in Fig.1.a

The CHOOZ oscillation limits (Fig.2) have been derived using an absolute method of 
analysis. All available experimental information has been compared to the expected 
no-oscillation values. The results of the analysis directly depend on the correct determination of the 
reactor power, absolute value of the ${\bar{\nu}}_e$ flux and their energy spectrum, nuclear fuel burn up 
effects, knowledge of the neutrino detection efficiency, absolute number of protons in the 
neutrino target and detector spectral response characteristics.

\vspace{5mm}

We would like to note here that CHOOZ experiment has demonstrated a revolutionary 
improvement on the neutrino detection technique: The level of the background at CHOOZ is 
hundreds times lower than has ever been achieved in any of the previous neutrino experiments at 
reactors. The underground position of the detector has sufficiently reduced the flux of cosmic 
muons, which is the main source of the correlated background and separation of the PM's from 
the neutrino target has reduced the accidentals coming from the radioactivity of the PM glass. 

\section{FUTURE KRASNOYARSK TWO DETECTOR EXPERIMENT Kr2Det}
\subsection{Main features of the approach}
\vspace{3mm}
(i) Relative to CHOOZ, we plan to increase the statistics of detected neutrinos by a factor of $\sim$ 20: 
$N_{{\nu}Kr2Det} \sim 20 N_{{\nu}CHOOZ}$.
To achieve this we increase the mass of the liquid scintillator target up to 50 ton.
(ii) To retain CHOOZ good effect to background ratio the detector is placed in an underground 
position. The overburden at Krasnoyarsk  is 600 m.w.e., which is twice as much as at the 
CHOOZ laboratory. To suppress the external gammas we chose a miniature version of the 
KAMLAND detector design with a $\sim$ 1 m thick layer of no-scintillating oil between the PM's and 
target volume (Fig.3).
 (iii) To eliminate most of the systematic uncertainties we turn to the idea of purely relative 
measurements and consider two identical scintillation spectrometers (far and near) stationed 
at 1100 and 250 m from the reactor.
\quad (iiii) To control remaining systematics special detector intercomparison procedures are being 
developed.
In Table 1 are presented some of (expected) parameters of the Kr2Det and of the CHOOZ 
experiments.

\begin{table*}[hbt]
\setlength{\tabcolsep}{1.5pc}
\newlength{\digitwidth} \settowidth{\digitwidth}{\rm 0}
\catcode`?=\active \def?{\kern\digitwidth}
\caption{Kr2Det $\&$ CHOOZ, neutrino detection rates and backgrounds}
\label{tab:effluents}
\begin{tabular*}{\textwidth}{@{}l@{\extracolsep{\fill}}ccccc}
\hline
Parameter & Distance (m) & M.W.E. & Target mass (ton) & ${\bar{\nu}}_{e} (d^{-1}$) &BKG ($d^{-1}$) \\
\hline
Kr2Det  Far & 1100 &  600    &  50 & 50     &   5  \\
Kr2Det Near &  250 &  600    &  50 & 1000   &   5  \\
CHOOZ       & 1100 &  300    &   5 & 12     &  1.6 \\
\hline
\end{tabular*}
\end{table*}
                                                                    
With a rate of 50 ${\bar{\nu}}_{e}$/day the desired total 40-50 thousand neutrino events can be accumulated in 
an acceptable time of data taking.

\vspace{5mm}

\subsection{Data analysis and detector calibrations}
\vspace{3mm}

In no-oscillation case the ratio of the two simultaneously measured positron energy 
spectra $S_{FAR}/S_{NEAR}$ is energy independent (Fig. 1b). Small deviations from the constant value 
of this ratio 
\begin{eqnarray}
S_{FAR}/S_{NEAR} = C(1 - \sin^{2}2{\theta}\sin^{2}{\phi}_{F})\times \nonumber \\
(1 - \sin^{2}2{\theta}\sin^{2}{\phi}_{N})^{-1}
\end{eqnarray}
are searched for oscillations (${\phi}_{F,N}$  stands for $1.27 {\Delta}m^{2}L_{F,N}/E_{\nu}$, 
and $L_{F,N}$ are the distances 
between the reactor and detectors). The results of the analysis are independent of the exact 
knowledge of neutrino flux and their energy spectrum, burn up effects, the numbers of target 
protons, neutrino detection efficiencies: \ However the relative difference of the detector energy 
scales should be strictly controlled. 

\vspace{5mm}

Calibration of the detectors is a key problem of the experiment considered here. \ Small 
difference of the response functions of the two detectors, which is difficult to avoid, can distort 
the ratio (4) and mimic (or conceal) the oscillation effect. \ The goal of the calibration procedures 
we consider is to measure sensitively this difference and introduce necessary corrections. This 
can be done by a combination of different methods. First we consider periodic control of the 
energy scales in many points using $\gamma$-sources shown in Fig. 4. \ Second method uses a small 
$^{252}$Cf spontaneous fission source placed in the detector's centers. The source generates continuous 
spectrum due to neutron recoils and prompt fission gammas also shown in Fig. 4. Any deviation 
from unity of the ratio of the spectra measured in two detectors can be used to calculate 
corrections under consideration. A useful monitoring of the scales \ at 2.23 MeV can also provide 
neutrons produced by through going muons captured by scintillator protons during the veto 
time.

\subsection{Expected constraints}

Expected 90\% CL oscillation limits can be seen in Fig.2. It was assumed that 40 000 ${\bar{\nu}}_{e}$ 
are detected in the detector stationed at \ 1100 m \ from the reactor and that the detector 
spectrometric difference \ is controlled down \ to 0.5\%.

\section{CONCLUSIONS}

Mass structure of the electron neutrino can sensitively be explored using two detector 
techniques. The results will also give useful normalization for future long baseline oscillation 
experiments at accelerators. 
The Kr2Det experiment is relatively inexpensive when compared to already running projects 
such as KAMLAND where neutrino target of a 1000 ton mass is used.

\section{Acknowledgements}

I would like to thank Professor McDonald for hospitality. I appreciate valuable discussions with 
E. Akhmedov, S. Bilenky, A. Piepke, and P. Vogel. 
Discussions with my colleagues in Kurchatov Institute and in particular with 
Yu. Kozlov and V. Sinev were of great help in 
preparing this report. This study is supported by RFBR grants NN 00-02-16035, 00-15-96708.

\section*{FIGURES}
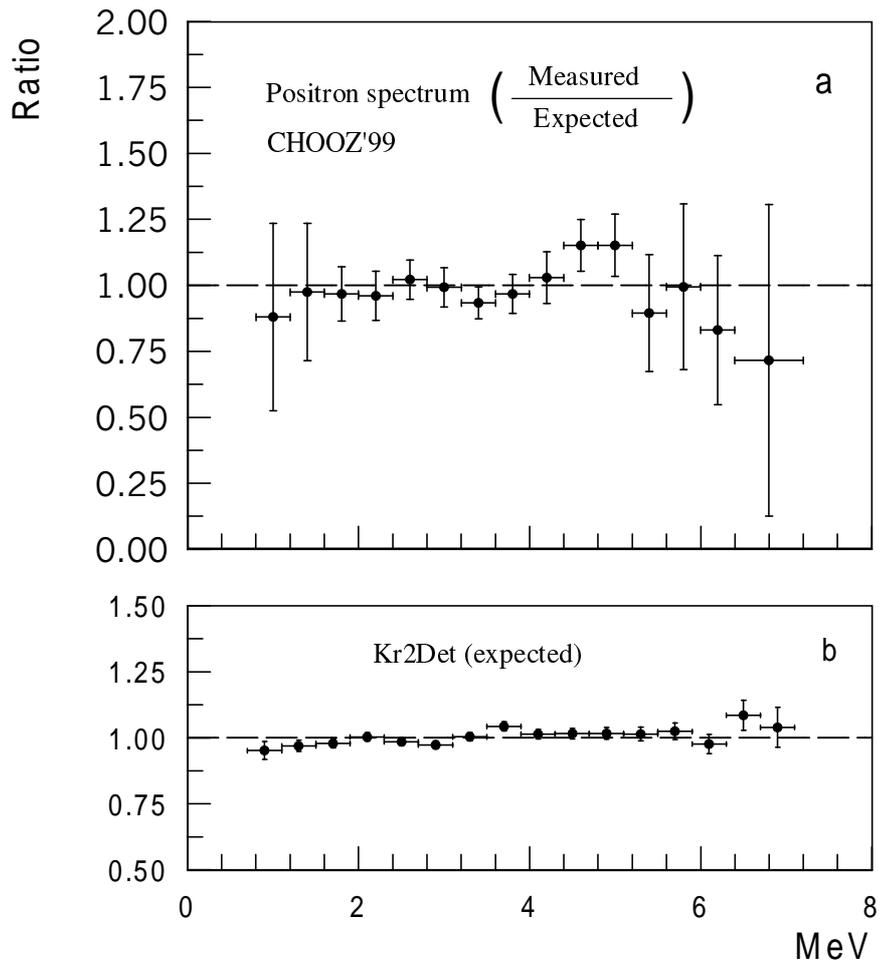
Figure 1. Ratios of positron spectra.\\
a) CHOOZ: Ratio of measured to expected in no-oscillation case.\\
b) Kr2Det : far to near (MC simulation, no-oscillations, normalized to unit).\\

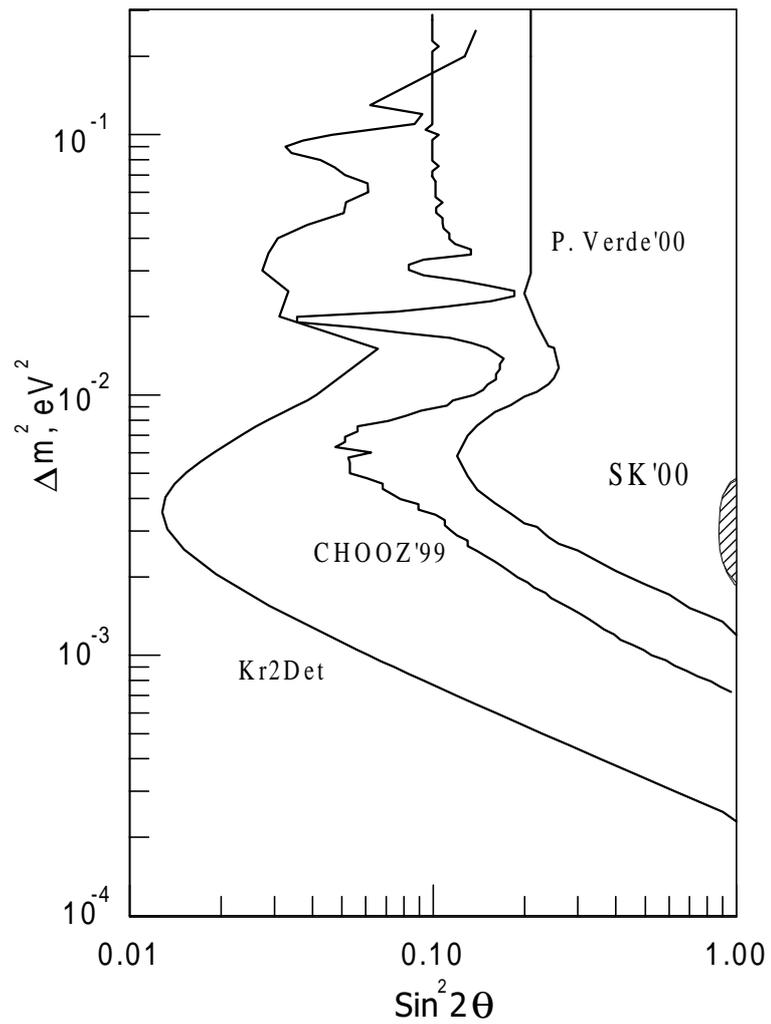
Figure 2. Reactor neutrino oscillation parameter plots.
"CHOOZ'99", "Palo-Verde 2000" and "Kr2Det" 
(expected)are 90\% CL ${\bar{\nu}}_{e}$ disappearance limits.
The shaded area is Super Kamiokande 2000 allowed ${\nu}_{\mu}\rightarrow {\nu}_{\tau}$ 
oscillation region.\\

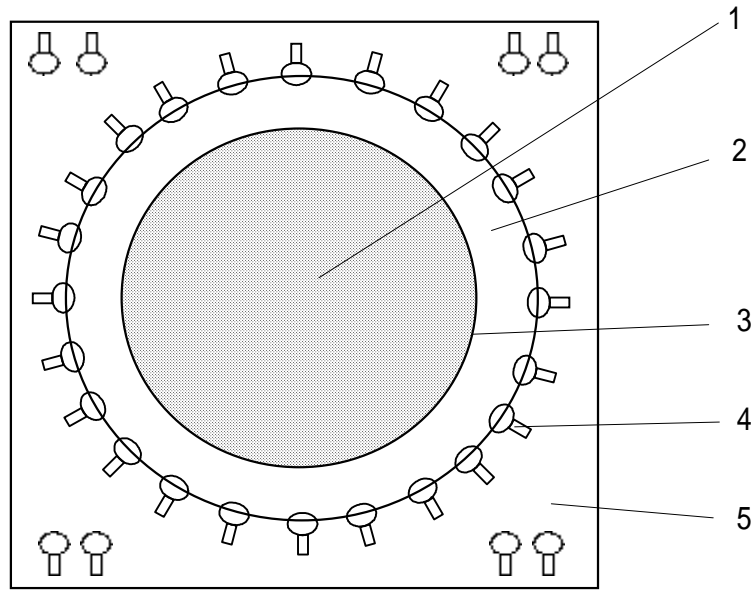
Figure 3. The Kr2Det detector. \\
1- Neutrino target, 50 ton mineral oil (PPO),\\
2 - mineral oil (buffer),\\
3 - transparent film, \\
4 - PMTs, 5 - veto zone.\\

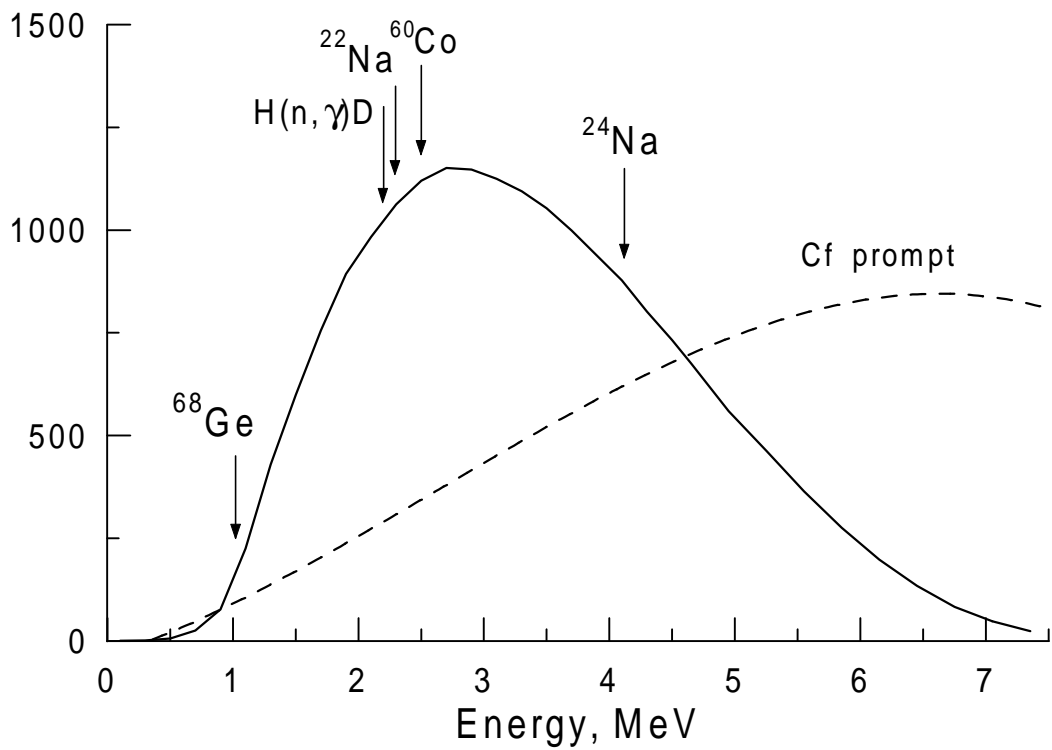
Figure 4. Sources for detector calibrations.\\
The solid line is the positron energy spectrum, 
the dashed line is the spectrum generated by Cf source.

\end{document}